# THE LINEAR BARYON

D. Bukta, G. Karl and B.G. Nickel

Department of Physics

University of Guelph

Guelph, ON N1G 2W1

Abstract

We describe classical orbits and quantum states of a three particle system, which is weakly chaotic. The quantum energies of the lowest 48 states which are invariant under permutations are given both for various approximations and to high accuracy. The quantum mechanical probability distributions are given in a two dimensional version of the system, the hexagonal pit. The quantum probabilities show strong resemblance to classical orbits of various types, including chaotic ones.

1.  INTRODUCTION

The simplest case of a three body system consists of three particles moving along a line under mutual interactions. With harmonic forces, the system is integrable trivially. With constant tension strings between particles the system becomes much more complex. We discuss the classical and quantum mechanics of this system. The motivation for constant tension strings comes from QCD, the theory of forces between quarks, but the principal interest in studying this system lies in the "weak chaos" it exhibits [1]. In contrast to a weakly chaotic system such as Henon-Heiles [2,3], this model has a potential that is a homogeneous function of distance so that chaotic orbits are present at all classical energies. But there are also other interesting aspects of this system; it admits two equivalent descriptions, both of which are extremely simple. One is the already mentioned "linear baryon", three quarks moving on a straight line, though we do not imply that real baryons are linear. The same system, in the centre of mass frame, can be represented by a single particle moving in the plane in a potential of hexagonal symmetry - we call this a "hexagonal pit" or hexagonal funnel. A particularly nice feature of the hexagonal representation is that it leads to very beautiful quantum probability distributions especially when plotted in color, as the reader may appreciate. Our discussion here is introductory, as we have more material available, which we hope to describe later [4]. We thought that the nice graphs are an appropriate tribute to Boris Stoicheff, whose long career in Physics always combined the interesting with the beautiful.

The hexagonal pit is a special case of "wedge billiards in a gravitational field" whose classical motion has already been studied [5]. The connection to a three body system is also known, though in an astronomical context of gravitational interaction between sheets of galaxies [6]. A related system of two bouncing balls of different masses falling in a gravitational field was studied by Whelan et al. [7]. The equal mass linear baryon corresponds to a wedge of half-angle $\pi/6$ or a ratio of three for the mass of the upper ball to that of the lower ball. It is known that this system has quasi-periodic and chaotic orbits coexisting in phase space. The quantum mechanics of the system has not been described as yet, so we stress more these aspects, though to make the paper self-contained we cover also classical trajectories which are described in a full hexagonal domain, not only in a single wedge. Our main interest lies in the transition from quasi-periodic motion to chaos, as seen through the quantum probabilities.

We discuss briefly the two representations of the system, a few selected classical orbits,

and two sets of variational wavefunctions which can label most quantum states. We give approximate and accurate estimates of energy levels, and then display probability distributions for some simple quantum states. We restrict the discussion to states which are completely symmetric under the hexagonal group. There are five other symmetry classes which we ignore in this preliminary account.

II.     SYSTEM DESCRIPTION

IIA.    Linear Baryon-three quarks on a line

The three quarks of equal mass m (to be set to unity later) have coordinates $x_1(t)$, $x_2(t)$ and $x_3(t)$. The three quarks are tied together by two strings of constant tension k (to be set equal to unity as well), so that the Hamiltonian is:

$$H = \left(p_1^2 + p_2^2 + p_3^2\right)/(2m) + (k/2)\left(|x_1 - x_2| + |x_2 - x_3| + |x_3 - x_1|\right) \quad [2.1]$$

where we have symmetrized the potential energy and divided by two; as noted there are only two strings tying the middle quark to the other two quarks. Which quark is in the middle varies as a function of time - the quarks pass through each other without hindrance, as they only feel the strings. The middle quark feels no net force and moves with constant velocity while the two outer quarks move with constant accelerations (unit for m=k=1) which pull them together. As the quarks pass through each other they change their ordering along the line. There are six orderings: $x_1 > x_2 > x_3$, $x_1 > x_3 > x_2$, ... and we assume that the centre of mass is at origin and at rest: $x_1 + x_2 + x_3 = 0$ and $p_1 + p_2 + p_3 = 0$. If these conditions hold initially the Hamiltonian [2.1] will maintain them later. Therefore the system has two degrees of freedom which can be taken as the Jacobi coordinates $\rho$ and $\lambda$:

$$x_1 - x_2 = \rho\sqrt{2}, \quad x_1 + x_2 - 2x_3 = \lambda\sqrt{6} \qquad [2.2]$$

in terms of which the internal Hamiltonian (still denoted by H) is

$$H = \left(p_\rho^2 + p_\lambda^2\right)/2 + \left(|\sqrt{3}\lambda - \rho| + |\sqrt{3}\lambda + \rho| + 2|\rho|\right)/\sqrt{8} \qquad [2.3]$$

where we have now set k=m=1.

IIB. <u>Hexagonal pit-one particle in the plane, and a description of orbits</u>

By considering $\rho$ and $\lambda$ as cartesian coordinates of a particle moving in the plane we can rewrite the Hamiltonian H most simply by using polar coordinates R, $\theta$ defined by

$$\lambda = R\cos\theta, \quad \rho = R\sin\theta. \qquad [2.4]$$

This is a maneuver similar to the use of hyperspherical coodinates [8] and we obtain from [2.3],

$$H = \left(p_R^2 + p_\theta^2/R^2\right)/2 + \left(\frac{R}{\sqrt{2}}\right)\left(|\sin\theta| + |\sin(\theta + 2\pi/3)| + |\sin(\theta - 2\pi/3)|\right). \qquad [2.5]$$

This is the Hamiltonian of the hexagonal pit. The equipotential lines of this Hamiltonian are regular hexagons. The orbit of the particle can be represented by a curve in the plane. Since this system is not harmonic there are no normal modes, but an infinity of trajectories. Most of these can nevertheless be described simply in terms of two prototype trajectories, having high and low "angular momentum" respectively. We use quotation marks since angular momentum is not conserved in a hexagonal potential. Although the average angular momentum of a closed trajectory is still meaningful, the use of this terminology is likely to lead to confusion and we will adopt a more descriptive terminology below. The high "angular momentum" mode is composed of segments of parabolas joined together to look almost like a circle (see trajectory A in Figure 1), and consequently we will describe such an orbit as whispering gallery. The low "angular momentum" trajectory consists of the particle moving along a groove of the hexagonal pit, corresponding to a diagonal of the hexagon, across to the opposite vertex and back, and will be called a channel mode. Each of these prototype trajectories can sustain transverse oscillations, perpendicular to the principal direction of motion as illustrated in Figure 2a. This is a rough description of most trajectories of the system.

When these trajectories are studied in a surface of section most of the area of phase space is occupied by such quasi-periodic trajectories, but a small area, about 5% of the total, corresponds to chaotic orbits[*]. This happens mostly in the region where the transverse excursions of the whispering gallery orbits approach the origin and/or the transverse excursion of the channel orbits allow overflow into a neighboring channel. The resulting orbits are quite irregular as illustrated in Figure 2b. An interesting feature of the chaotic orbit in Figure 2b is that it does not fill uniformly the whole hexagon. The probability distribution for this orbit shows high and low probability areas with hexagonal symmetry. Note the high probability areas are in the wrong places for a quasiperiodic orbit of either prototype, they are in the centre of the hexagon and on the ridges between the grooves, but are necessary to guarantee the average over all modes covers the plane uniformly.

Each orbit in the hexagonal pit corresponds to an orbit of the linear baryon. The whispering gallery orbit corresponds to a rapid permutation over all orderings of the three quarks, while the channel orbit corresponds to a quark oscillating against a "diquark" (two quarks joined together with zero distance between them). Figure 3 gives the linear baryon counterparts to three trajectories A, B and C of the hexagonal pit illustrated in Figure 1. Our description of classical orbits differs from that in the literature which is limited to motion in a single "half-wedge", a one-twelfth of the hexagon. We find the description of the whole orbit more intuitive and easier to follow, though of course the motion in the hexagon consists of segments in successive wedges.

---

[*]Our surface of section at $\rho = 0$ and energy $E = 1$ is that given in Figure 3 of reference [7] with the assignments $v_2 = 2p_\lambda/\sqrt{3}$, $x_2 = 4\lambda/\sqrt{6}$.

### III. QUANTUM MECHANICAL STATES AND QUANTUM NUMBERS

The Hamiltonian of the system may be considered as a quantum operator and, without loss of generality, we can take $\hbar = 1$ since together with the mass m and tension k it sets the scale for the problem. We found most useful the hexagonal pit representation [2.5] to describe the motion. The corresponding Schrödinger equation is two dimensional in R and θ, but of course it is not separable since it lacks rotational symmetry. To describe the symmetry classes of the solutions, the variational approximation is helpful (for all states, see ref. [9]), using two dimensional harmonic oscillator states as trial wave functions, with the frequency of the oscillator as variational parameter (separately for each state). The two dimensional oscillator wavefunctions have two quantum numbers: a radial quantum number "n" and an azimuthal quantum number "m". The quantum number m governs the angular dependence of the wavefunction through trigonometric functions cos(mθ) or sin(mθ), with m = 0, 1, 2, ...; in this case the complex representation is less useful. The complete wavefunction has the form

$$\psi_{nm}(\omega, R, \theta) = N \begin{pmatrix} \cos(m\theta) \\ \sin(m\theta) \end{pmatrix} (\omega R^2)^{m/2} \cdot \exp(-\omega R^2/2) \cdot L_{nm}(\omega R^2) \qquad [3.1]$$

where $L_{nm}(x)$ are Laguerre polynomials [10]. The factor N normalizes the wavefunction squared with $RdRd\theta$. This set of wavefunctions is orthonormal and complete at any fixed ω. One can express the Hamiltonian H as a matrix in a basis provided by this set of wavefunctions, and diagonalize it. As a first approximation we minimize diagonal matrix elements with respect to ω to get an estimate of the energy of the state, which is compared to a much more accurate determination obtained by a numerical solution of the Schrödinger equation. It is important to note that the quantum number m is only "good" modulo six for a hexagonal potential; the potential energy in [2.5] will connect states with m = 0, 6, 12, 18, ... similarly for states with m = 1, (-)5, 7, (-)11, 13, ..., etc. Nevertheless we can still use these higher values to label states of the hexagonal pit, if their expansion in terms of harmonic oscillator states has a leading term in $\cos(m\theta)$, say, with many other subleading components. Therefore we shall label the states by n,m with m often large.

The states with m = 1, 2, 3, 4, 5 correspond to different symmetry classes of the hexagonal group, and as noted in the introduction we are only discussing in this paper states with full hexagonal symmetry (corresponding to the representation A1 of the hexagonal group D3). The symmetry group of the hexagonal pit is called D3, and this group has six irreducible representations, usually labelled by A1, A2, B1, B2, E1, E2. The representations E1 and E2 are each two-dimensional, the others are one dimensional. A1 is the representation which is totally symmetric under all transformations of the hexagon. The wavefunctions belonging to A1 are linear combinations of $\cos(m\theta)$ factors with m = 0, 6, 12, ... and are the only ones we shall consider here. Other values of m (modulo six) correspond to other representations. Each of the above symmetry classes has a counterpart in the linear baryon representation, but we shall not dwell on this here. The class A1 corresponds to wavefunctions which are symmetric under permutations of the three quarks and even under parity.

Table I gives the values of the variational energies for eigenstates, together with the "exact" values obtained by numerical solution of the coupled Schrödinger equations in R,θ. The "exact" values are uncertain only in the last digit quoted; these values have been checked by diagonalizing the Hamiltonian in large (~100) harmonic oscillator bases with ω variable and minimizing each eigenvalue separately. The oscillator variational estimates, based on diagonal elements only, are better than 1% for n<<m but deteriorate to the point of becoming meaningless for n>>m. This is as expected since a basis in states of circular symmetry should describe well whispering gallery orbits but not channel states. More than this cannot be said purely on the basis of the elementary quantum variational calculation but we have also studied the problem in quasiclassical approximation. By a careful numerical evaluation of the angular and radial actions of classical quasiperiodic whispering gallery modes we conclude that one should restrict the harmonic oscillator quantum numbers by n≲m/2.

For channel states we make the almost trivial observation that the hexagonal symmetry is irrelevant and replacing the potential by a square pyramidal pit would not dramatically alter the orbits through the origin. Now the square pyramidal pit is a known separable problem, with conserved energy parallel and transverse to the groove and so a much better approximation is obtained by using as trial wavefunctions products of Airy wavefunctions along a single diameter and perpendicular to it, with scale factors in the arguments of the two Airy functions as two variational parameters. The corresponding energies are given also in Table I and it can be seen that especially at high quantum number n for m = 0, this is a very good approximation. Here too we have investigated the problem quasi-classically and conclude that the restriction on the parallel and transverse Airy quantum numbers $n_\lambda$, $n_\rho$ is $n_\lambda \geq 4n_\rho$.

There is a region of overlapping validity of the two approximations and this enables us to identify

$$m = 3n_\rho \quad , \quad n_\rho + n_\lambda = 2n + m ; \qquad [3.2]$$

both sets of labels are given in Table I. These quantum numbers for the variational approximations can, without ambiguity in most cases, be also used for the exact states on the basis of the near equality of the energy eigenvalue and/or the eigenfunction. Exceptions are noted in Table 1. The pair of entries marked (**) are mixed states with $\Delta n_\rho = 2$ and $\Delta n_\lambda = 4$ and can be identified, essentially as described by DeLeon, et al. [3], with the 2:1 resonance orbits shown in Figure 1 as C and D. The mixed pair marked (*) are not reasonably identified with any classical orbit since one of the partners of the resonance comes from a channel state and the other from whispering gallery. The classical orbits exist in disjoint regions of phase space and thus the mixing is best viewed as a purely quantum tunnelling phenomena.

Although it is possible to describe states in the region of overlapping validity of the two variational approximations by either set of quantum numbers n,m or $n_\rho,n_\lambda$ related in equation [3.2], there is another description that more accurately describes the physics in this region. Specifically, this region is the classically chaotic region associated with the unstable periodic "bouncing ball" orbit in which the particle moves in a straight line through the origin from ridge to ridge; the most appropriate description of the quantum states reflects this influence without reference to the quasi-periodic orbits on either side. Now the role of unstable orbits in Gutzwiller's trace formula for systems displaying hard chaos is well known but we believe that it has not been adequately appreciated that the 1-dimensional quantization that is the basis of the trace formula is even more appropriate in the case of weak chaos as studied here.

Let $S = \int p \cdot dq$ be the classical action on the unstable orbit. In our case it is easily determined from the one dimensional Hamiltonian $p_\rho^2/2 + \sqrt{2}\rho$ that describes the "bouncing ball" motion as $S = 8 E_{res}^{3/2}/3$. Quantization of this action including the topologically invariant Maslov index for the unstable orbit [11] gives

$$E_{res} = \left(3\pi(n_\rho + n_\lambda + 1)/4\right)^{2/3} = \left(3\pi(2n + m + 1)/4\right)^{2/3} . \qquad [3.3]$$

Equation [3.3] is not meant to imply an energy level for every combination of n,m or $n_\rho, n_\lambda$ but rather that if there is a state with an integer combination that lies within or close to the classically chaotic band (which we have estimated as $n \lesssim m/2$ and $n_\lambda \gtrsim 4n_\rho$) then the energy of that state must be given by [3.3]. We purposely leave "close to" vague; how to quantify this is presently under investigation. Several comments however are in order. First, it has often been observed in the literature that in systems of hard chaos the influence of unstable orbits of low order is surprisingly strong. Here too it is to be noted that the estimates [3.3] given in Table 1 for the chaotic band are all better (or at least as good) as those from the variational calculations. Further evidence for the appropriateness of our prescription comes from the wavefunctions described in section IV. Second, there will almost certainly be situations in which the chaotic region is so small that no states deduced from a formula like [3.3] lie within the chaotic region. In that case the more reasonable prescription for semiclassical quantization is likely to be the combination of "smoothing" and standard EBK analysis suggested by Reinhardt and Jaffe [12].

## IV. PROBABILITY DISTRIBUTIONS

Figure 4 gives some typical probability distributions of quantum states which have been chosen at approximately the same energy but all possible quantum numbers m to illustrate the correspondence to the prototypical classical orbits discussed in section II. The colors correspond to different probabilities, the linear scale is indicated. The first distribution corresponds to a channel mode, in which the particle moves along a groove of the hexagonal pit (n=12, m=0). The quantum state is a linear superposition of motions along the three diagonals of the hexagon, to realize the A1 symmetry class. The probability is highest at the turning points (where the velocity vanishes) and in the center where there is a nice constructive interference between the three beams. The probability in the centre is estimated to be about ten times ($3^2$) the value midway between the turning point and the centre. The zeros of the radial wavefunction can be counted on the picture and correspond to n=12. It should be emphasized that this is an accurate numerical wavefunction which has quite a bit of $\cos(m\theta)$ components with m = 6, 12, 18, ...48 to account for the small wavefunction ("zero" on the picture) at large radii, between the separate beams.

The last distribution on Figure 4 shows a probability distribution for a state which corresponds to a whispering gallery orbit (n = 2, m = 24). The circular type motion is clearly visible, as well as the transverse (radial) excitation, with the two radial nodes. The maximum probability occurs halfway between grooves where the velocity is smallest but this is a small difference. The two distributions in the middle correspond to states which have both radial and angular excitation and the probability has large values throughout the hexagon.

The state with n=5 and m=12 lies approximately in the band of chaotic states and its probability distribution is resemblant of the chaotic classical orbit of Figure 2b with large probabilities in the centre and midway between the grooves. This is shown in more detail in Figure 5 which compares the classical probability distribution corresponding to the chaotic orbit of Figure 2b with the quantum probability of Figure 4, (n5m12). The near perfect correspondence of probabilities in Figure 5 leaves little doubt that n=5, m=12 is correctly assigned as chaotic. Nevertheless the quantum probability distribution is very nicely regular, and it would be hard to guess that it corresponds to a chaotic region. De Leon, et al [3] have emphasized that complicated nodal patterns are not necessarily an indicator of chaos; the state n=5, m=12 is a nice illustration of the converse, namely that a regular nodal pattern does not necessarily imply the absence of chaos.

## V. CONCLUSION AND OPEN ISSUES

We have discussed in detail the quantum states of the hexagonal pit which belong to the class A1. Their relation to the corresponding classical trajectories is quite transparent in most cases, including some chaotic ("ergodic") orbits. We have emphasized the role of 1-dimensional quantization on unstable periodic orbits in describing the chaotic states.

The generalization of the results presented here to other symmetry classes of the hexagonal pit is one of the first extensions that suggests itself. Less immediately, the generalizations to other pits, corresponding to other regular polygons should be attempted. The triangular pit has been studied by Szeredi and Goodings [13] and is a nice example of hard chaos. The square pit is integrable while the pentagonal pit may be of particular interest as an example of hard chaos meeting near integrability on a very sharp boundary. The whispering gallery region resonances are extremely narrow with chaos in this region almost undetectable. On the other hand the symmetry of the pentagon precludes stable channel orbits and results in a region of phase space of hard chaos. This would make it a useful system to study further to help us understand better the quantum mechanical edges of chaos.

VI. ACKNOWLEDGMENTS

This work was supported by NSERC grants.

FIGURE CAPTIONS:

Figure 1.       Examples of trajectories in the hexagonal pit.  Trajectory A is the whispering gallery orbit, while B, C, D are transverse excitations of the channel orbit. Trajectory C(D) is a 2:1 resonance which is stable (unstable).

Figure 2a.      Examples of trajectories which have larger transverse excitations of the basic whispering gallery and channel orbits than shown in Figure 1.

Figure 2b.      Example of the start of a chaotic orbit.

Figure 3.       The linear baryon counterparts of the three trajectories A, B, C of the hexagonal pit shown in Figure 1.  The trajectories x(i) of the three quarks as a function of time (t) are given.

Figure 4.       The quantum probability distributions for four excited states of the hexagonal pit. The notation n12m0 means the state n=12, m=0.  The color scale of relative probabilities is given at the bottom.  The graphs have been rescaled to make the hexagons, the classical boundaries of the orbits for the given E, the same size.

Figure 5a.      The classical probability distribution for the chaotic orbit of Figure 2b estimated by a (long) time average of the position.

Figure 5b.      The quantum probability distribution of the state n=5, m=12 which is also displayed in color Figure 4.

**Table 1.** Comparison of energies for states of $A_1$ symmetry as obtained by the indicated calculation methods. Exact values that are marked * (**) are mixed states and hence have an ambiguous labelling.

| n | m | Exact | H.O. var'n | 1-d Quasi-classical (2n+m+1) | Airy var'n | $n_8$ | $n_D$ |
|---|---|---|---|---|---|---|---|
| 0 | 0 | 1.6843724 | 1.69088 | | 1.68729 | 0 | 0 |
| 1 | 0 | 3.5559970 | 3.54142 | | 3.62667 | 2 | 0 |
| 2 | 0 | 5.0003994 | 4.98763 | | 5.04238 | 4 | 0 |
| 3 | 0 | 6.2365647 | 6.24630 | | 6.25752 | 6 | 0 |
| 4 | 0 | 7.3424864 | 7.38816 | | 7.35290 | 8 | 0 |
| 5 | 0 | 8.3595396 | 8.44737 | | 8.36522 | 10 | 0 |
| 6 | 0 | 9.3116203 | 9.44370 | | 9.31510 | 12 | 0 |
| 7 | 0 | 10.213241 | 10.3899 | | 10.21557 | 14 | 0 |
| 8 | 0 | 11.073906 | 11.2946 | | 11.07554 | 16 | 0 |
| 9 | 0 | 11.900291 | 12.1645 | | 11.90145 | 18 | 0 |
| 10 | 0 | *12.697993 | 13.0042 | | 12.69812 | 20 | 0 |
| 11 | 0 | 13.468880 | 13.8177 | | 13.46931 | 22 | 0 |
| 12 | 0 | 14.21802 | 14.6078 | | 14.21798 | 24 | 0 |
| 13 | 0 | 14.94758 | 15.3771 | | 14.94655 | 26 | 0 |
| 14 | 0 | **15.64786 | 16.1276 | | 15.65704 | 28 | 0 |
| 15 | 0 | 16.34869 | 16.8610 | | 16.35110 | 30 | 0 |
| 16 | 0 | 17.02871 | 17.5788 | | 17.03016 | 32 | 0 |
| 0 | 6 | 6.6225651 | 6.59604 | | 6.55520 | 4 | 2 |
| 1 | 6 | 7.7365445 | 7.67598 | | 7.74827 | 6 | 2 |
| 2 | 6 | 8.7894295 | 8.68812 | (11)8.7580 | 8.82999 | 8 | 2 |
| 3 | 6 | 9.7878106 | 9.64723 | (13)9.7897 | 9.83294 | 10 | 2 |
| 4 | 6 | 10.737520 | 10.5630 | | 10.77595 | 12 | 2 |
| 5 | 6 | 11.643140 | 11.4425 | | 11.67114 | 14 | 2 |
| 6 | 6 | 12.508507 | 12.2907 | | 12.52692 | 16 | 2 |
| 7 | 6 | 13.337758 | 13.1118 | | 13.34940 | 18 | 2 |
| 8 | 6 | 14.13553 | 13.9087 | | 14.14320 | 20 | 2 |
| 9 | 6 | 14.90591 | 14.6842 | | 14.91194 | 22 | 2 |
| 10 | 6 | **15.66427 | 15.4404 | | 15.65851 | 24 | 2 |
| 11 | 6 | 16.38527 | 16.1790 | | 16.38525 | 26 | 2 |
| 12 | 6 | 17.09389 | 16.9015 | | 17.09412 | 28 | 2 |
| 0 | 12 | 10.067030 | 10.0560 | | 9.91975 | 8 | 4 |
| 1 | 12 | 10.968205 | 10.9491 | | 10.90977 | 10 | 4 |
| 2 | 12 | 11.843901 | 11.8088 | | 11.84340 | 12 | 4 |
| 3 | 12 | *12.697001 | 12.6398 | | 12.73139 | 14 | 4 |
| 4 | 12 | 13.531453 | 13.4457 | (21)13.4778 | 13.58142 | 16 | 4 |
| 5 | 12 | 14.34564 | 14.2293 | (23)14.3205 | 14.39919 | 18 | 4 |
| 6 | 12 | 15.13987 | 14.9930 | (25)15.1391 | 15.18906 | 20 | 4 |
| 7 | 12 | 15.91375 | 15.7388 | | 15.95445 | 22 | 4 |
| 8 | 12 | 16.66705 | 16.4681 | | 16.69814 | 24 | 4 |
| 0 | 18 | 12.996033 | 12.9854 | | 12.77984 | 12 | 6 |
| 1 | 18 | 13.790631 | 13.7770 | | 13.65922 | 14 | 6 |
| 2 | 18 | 14.56862 | 14.5473 | | 14.50252 | 16 | 6 |
| 3 | 18 | 15.33147 | 15.2987 | | 15.31480 | 18 | 6 |
| 4 | 18 | 16.08072 | 16.0330 | | 16.10010 | 20 | 6 |
| 5 | 18 | 16.81799 | 16.7518 | | 16.86160 | 22 | 6 |

```
0 24    15.62393   15.6124           15.34753   16  8
1 24    16.34971   16.3368           16.15354   18  8
2 24    17.06374   17.0462           16.93368   20  8
```

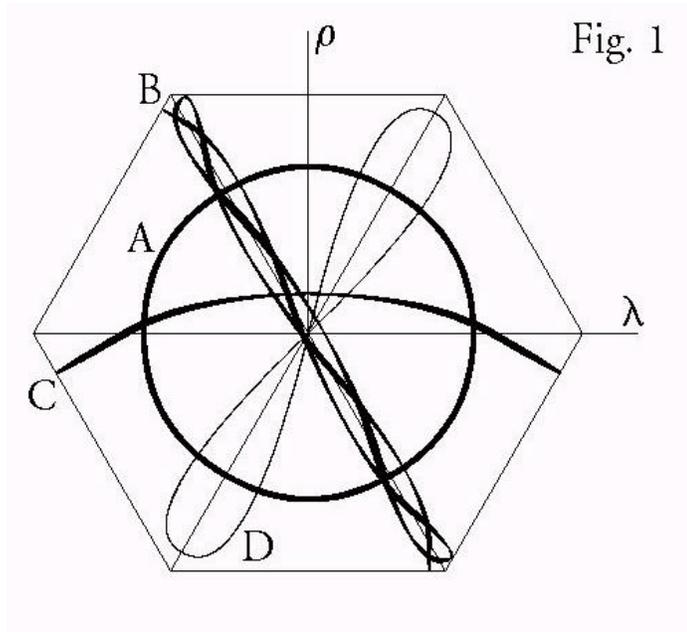

Fig. 1

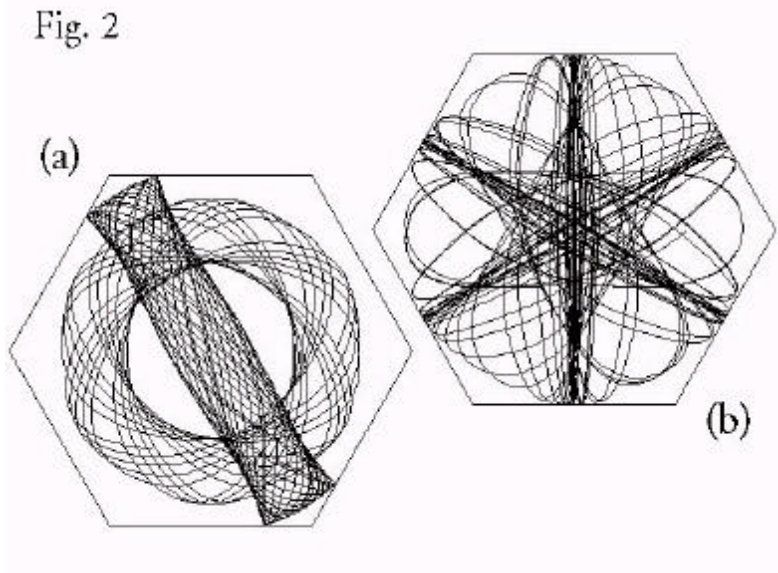

Fig. 2

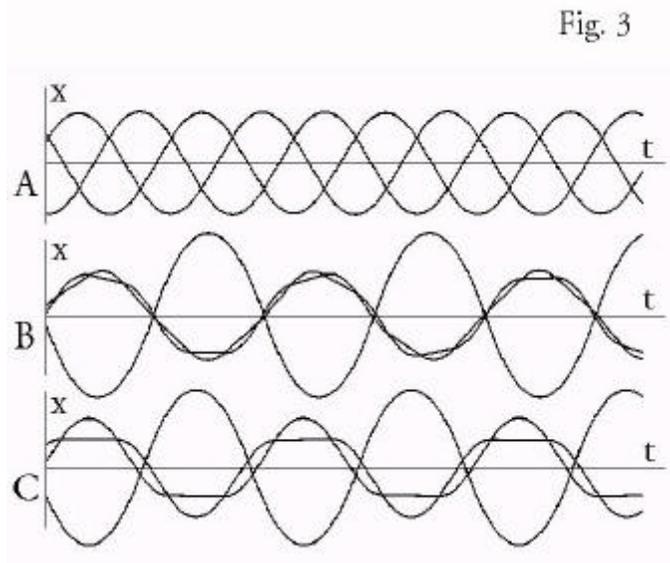

Fig. 3

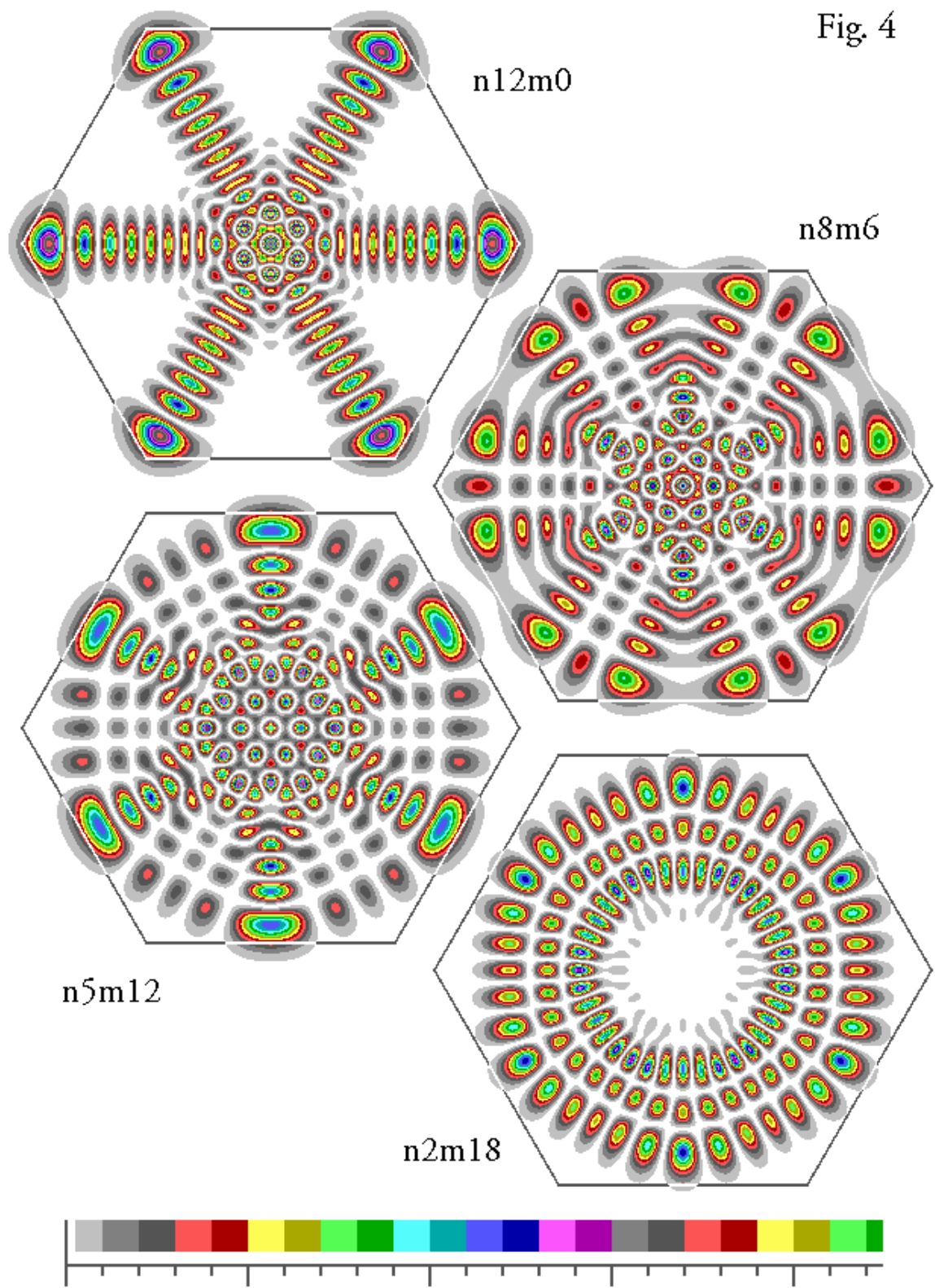

Fig. 4

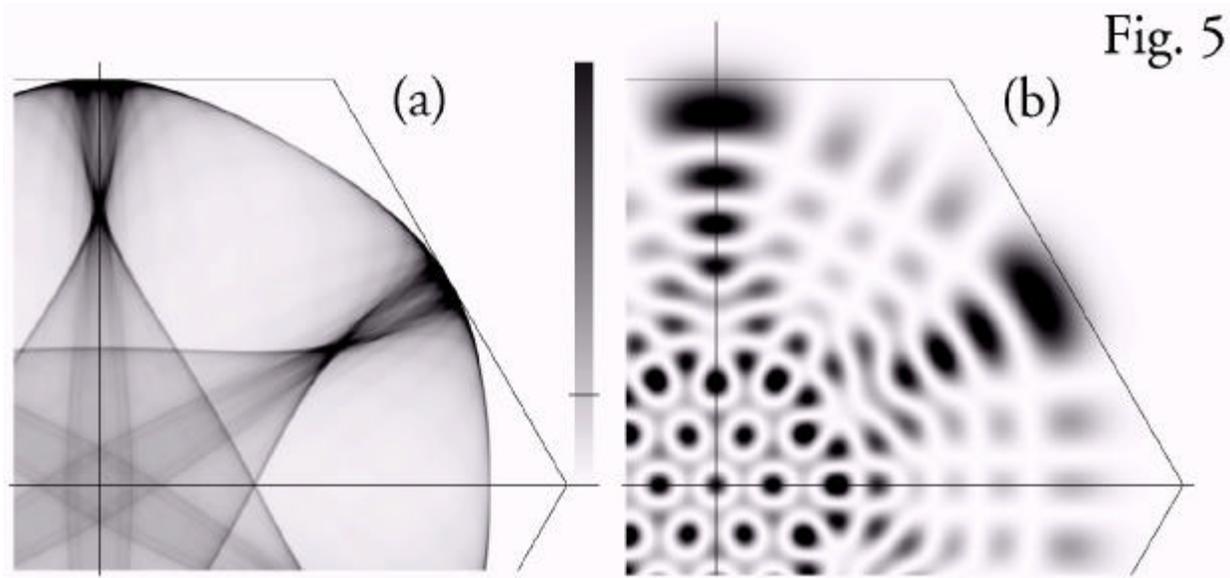

Fig. 5